\documentclass[preprint,12pt]{elsarticle}

\usepackage{amssymb}

\usepackage[percent]{overpic}
\usepackage{float}
\usepackage{pgfplots}
\usepackage{placeins}
\usepackage{amsmath}
\usepackage{amsthm}
\usepackage{color}
\usepackage{amssymb}
\usepackage{mathtools}
\usepackage{subfigure}
\usepackage{multirow}
\usepackage{epsfig}
\usepackage{listings}
\usepackage{enumitem}
\usepackage{rotating,tabularx}
\usepackage{graphicx}
\usepackage{graphics}
\usepackage{epstopdf}
\usepackage{longtable}
\usepackage{hyperref}
\usepackage{epigraph}
\usepackage{xspace}
\usepackage{amsfonts}
\usepackage{eurosym}
\usepackage{ulem}
\usepackage{footmisc}
\usepackage{comment}
\usepackage{setspace}
\usepackage{geometry}
\usepackage{caption}
\usepackage{pdflscape}
\usepackage{array}
\usepackage{booktabs}
\usepackage{dcolumn}
\usepackage{mathrsfs}

\newcommand{\bc}{\begin{center}}
\newcommand{\ec}{\end{center}}

\renewcommand{\ast}{{}^{\textstyle *}} 

\newcommand{\Sref}[1]{Section~\ref{sec:#1}}

\newcommand{\Cref}[1]{Corollary~\ref{coro:#1}}
\newcommand{\cref}[1]{Cor.~\ref{coro:#1}}

\newcommand{\eg}{{\it e.g.}\xspace}

\newcommand{\fref}[1]{Fig.~\ref{fig:#1}}
\newcommand{\Fref}[1]{Figure~\ref{fig:#1}}

\newcommand{\be}{\begin{equation}}
\newcommand{\ee}{\end{equation}}
\newcommand{\bea}{\begin{eqnarray}}
\newcommand{\eea}{\end{eqnarray}}
\newcommand{\bi}{\begin{itemize}}
\newcommand{\ei}{\end{itemize}}

\journal{Chaos, Solitons and Fractals}

\begin{document}

\begin{frontmatter}



\title{Measures of physical mixing evaluate the economic mobility of the typical individual}


\author[inst7,inst6]{Viktor Stojkoski\corref{cor1}}

 \cortext[cor1]{Corresponding author e-mail: vstojkoski@eccf.ukim.edu.mk}

 \affiliation[inst7]{organization={Faculty of Economics, Ss.~Cyril and Methodius University},
            city={Skopje},
            postcode={1000}, 
            country={North Macedonia}}   

\affiliation[inst6]{organization={Center for Collective Learning, ANITI, University of Toulouse},
            city={Toulouse},
            postcode={31000}, 
            country={France}}  

\begin{abstract}
Measures of economic mobility represent aggregate values for how individual wealth changes over time. As such, these measures may not describe the feasibility of a typical individual to change their wealth. To address this limitation, we introduce mixing, a concept from statistical physics, as a relevant phenomenon for quantifying how individuals move across the wealth distribution. We display the relationship between mixing and mobility both theoretically and using data. By studying the properties of an established model of wealth dynamics, we show that some individuals can move across the distribution when wealth is a non-mixing observable. Only in the mixing case every individual is able to move across the whole wealth distribution. There is also a direct equivalence between measures of mixing and the magnitude of the standard measures of economic mobility, but the opposite is not true. We then describe an empirical method for estimating the mixing properties of wealth dynamics in practice. We use this method to present a pedagogical application using the USA longitudinal data. This, approach, even though limited in data availability, leads to results suggesting that wealth in the USA is either non-mixing or that it takes a very long time for the individuals to mix within the distribution. These results showcase how mixing can be used in tandem with measures of mobility for drawing conclusions about the extent of mobility across the whole distribution.
\end{abstract}

\begin{highlights}
\item We introduce mixing, as a relevant phenomenon for quantifying the mobility of the typical individual.
\item We show that there is a direct equivalence between mixing and standard measures of mobility only when every individual is able to move across the wealth distribution.
\item We also describe a pedagogical empirical method for studying mixing in real-world economies, and demonstrate its application using the USA longitudinal data.
\item Mixing can be used in tandem with measures of mobility for drawing conclusions about the extent of mobility across the whole distribution.
\end{highlights}


\end{frontmatter}


\section{Introduction}\label{sec:introduction}
Economic mobility describes ``dynamic aspects of inequality'' \cite{Shorrocks1978}. It quantifies how wealth (or income) changes within a population over a given time period~\cite{markandya1984welfare}. Intuitively, when mobility is high, the chances of an individual to change their wealth (either in absolute or relative terms) over the time period are high. When mobility is low, individuals are unlikely to change their wealth over time, or such changes are slow.

Mobility measures are assumed to be derived from the joint distribution of wealth at two points in time. On this basis, Shorrocks \cite{Shorrocks1978} described several required properties for the statistical measurement of mobility and set the standard for such measures. A particular feature of measures of economic mobility is that they represent an aggregation of the observed changes in the individual wealth~\cite{cowell2018measuring}. This means that whenever there is a change in the wealth of \textit{any} individual wealth, the measures will suggest the existence of mobility. As a result, in certain circumstances mobility measures may not describe the feasibility of the \textit{typical} individual to change their wealth. 

Here we address this issue by introducing \textit{mixing} as a relevant concept when quantifying the feasibility of \textit{every} individual in the economy to change their wealth. Mixing is a well-known concept in statistical physics. It describes the property of a dynamical system being strongly intertwined. The idea is that in a mixing economy, if the wealth of a group describing the typical individuals is followed over time, the distribution of wealth within this group will gradually converge to the population wealth distribution. This means that any measure of mixing will evaluate the extent to which \textit{every} individual in an economy is able to move across the whole wealth distribution. When the convergence is rapid, we could interpret that as high wealth mixing. Long convergence times is interpreted as low mixing. If wealth dynamics are non-mixing, measures of mixing will suggest that there is no convergence. 

It is important to clarify that there is no strict empirical evidence to confirm that the wealth of any group of individuals will invariably converge to the population’s wealth distribution over time. However, the theoretical concept of ``mixing'' in an economy provides a valuable framework for exploring this potential convergence. This paper delves into the conditions and dynamics within an economy that could allow for the wealth of what we term as ‘a group describing the typical individuals in the economy’ to align with the overall wealth distribution. Conversely, we also explore the characteristics and economic scenarios that might inhibit this convergence. 

We introduce two measures of mixing: 1) the relaxation time and 2) the mixing time, and use them to investigate the relationship between mixing and mobility both theoretically and using empirical data. The relaxation time represents the time it takes for a wealth distribution to converge to a target distribution, essentially measuring the speed at which individuals mix into the overall wealth distribution. The mixing time in contrast, is defined as the time at which the wealth distribution of individuals first becomes statistically identical to a target distribution, offering a different perspective on the timescale of economic mixing. Both measures quantify the timescale over which individuals mix into the wealth distribution (lower values imply higher mixing), but have distinct estimation procedures. On the one hand, the relaxation time can be easily estimated in theoretical models by exploiting their spectral properties. This measure, however, is difficult to quantify in practice. On the other hand, the mixing time can be easily calculated using empirical data, but difficult to study analytically.

In this work, we present results about the mixing of wealth and compare them to intragenerational mobility. The focus on intragenerational mobility arises from the convenience of available data and the established analytical properties of mathematical models specifically designed for studying intragenerational wealth dynamics. The same ideas can be applied to income and/or intergenerational mobility, but for the purpose of this study, we use the terms ``mobility'' and ``intragenerational mobility'' interchangeably.

In particular, we use the relaxation time to study the theoretical properties of mixing in a model that is able to distinguish between mixing and non-mixing wealth dynamics, called Reallocating Geometric Brownian Motion (RGBM) \cite{MarsiliMaslovZhang1998,BouchaudMezard2000,LiuSerota2017,BermanPetersAdamou2019,kemp2022statistical,kemp2023learning}. RGBM is a null model of an exponentially-growing economy~\cite{stojkoski2020generalised,stojkoski2019cooperation,stojkoski2021evolution}. It has three parameters representing common economic growth, random shocks to individual wealth, and economic interaction among agents, quantified by a reallocation rate. The advantage of using RGBM is that the model distinguishes mixing and non-mixing regimes solely by the sign of the reallocation parameter. Besides this, the model is also known to reproduce several important stylized facts, such as the often-observed Pareto tail of the population wealth distribution, thus making it easily relatable to real-world wealth dynamics. We find that in the mixing regime of RGBM, the relaxation time scales with the inverse of the reallocation rate. Then there is also a direct relationship between the relaxation time and three standard measures of economic mobility: the Spearman rank correlation, the intragenerational earnings elasticity, and the wealth transition matrices. As the reallocation rate becomes higher, i.e., as a larger share of each individual's wealth is pooled and then shared per unit time, the relaxation time becomes shorter proportionally, implying that mixing occurs sooner and that mobility increases. As the reallocation rate approaches zero, relaxation times get longer, and mobility lower. Finally, when the reallocation rate is below zero, even though standard measures of economic mobility might suggest the existence of mobility, there is no mixing. Thus, if we use standard measures of mobility in a non-mixing system, we may conclude misleadingly that everyone is able to move across the wealth distribution. 

We use the mixing time to devise an empirical method for evaluating the wealth mixing within an economy. This empirical method is inherently flexible and not model dependent. It requires longitudinal data and tracks the time until the distribution of a subsample of the population representing the typical individuals resembles the wealth distribution of the whole economy. We showcase its application using survey data on the USA household wealth taken from the Panel Study of Income Dynamics (PSID) dataset~\cite{hill1992panel}. We find results which suggest that wealth is either non-mixing or that it takes a long time for the individuals to mix within the distribution. Nevertheless, the PSID dataset has several limitations that restrict its practical application and therefore the analysis presented here is only pedagogical. We believe that with the rapid development of data gathering methods and the improved understanding of wealth dynamics within a population, some of this shortcoming will be overcome, yielding a more in-depth interpretation of mixing in real economies. 

These results improve our understanding about the measurement of mobility by showcasing how physical measures of mixing can be used to track the mobility of the typical individual in the economy. As such, they can help policy makers in designing interventions that target the well-being of these individuals. More importantly, they can be used in tandem with standard measures of mixing for deriving a more holistic interpretation about the extent of mobility in the economy.

The paper is organized as follows. In~\Sref{lit-review} we discuss the approaches to the measurement of mobility and describe how our study about mixing can provide new valuable information. Then, in~\Sref{mixing-time} we define mixing mathematically and compare its characteristics to those of standard mobility measures. In the same section, we introduce the relaxation and mixing times as measures that evaluates the degree of mixing in an economy. \Sref{rgbm} relates mobility and mixing in reallocating geometric Brownian motion as a model of wealth. \Sref{empirical-results} presents the empirical method for evaluating the mixing state of an economy. We discuss our findings in~\Sref{discussion}.

\section{The measurement of economic mobility}
\label{sec:lit-review}

The literature has identified two approaches for measuring economic mobility: relative and absolute mobility~\cite{dardanoni1993measuring,fields1999measurement}. 

The relative approach measures economic mobility by comparing the ranks of individuals (or groups of individuals) between two time points. For instance, Prais~\cite{prais1955measuring} used ranks to measure the mobility between social classes in the United Kingdom , whereas White~\cite{white1963cause} used them to study the income dynamics in Denmark. Motivated by these analyses, Shorrocks~\cite{Shorrocks1978} laid down the theoretical groundwork for the properties of these measures. The author introduced axiomatic properties to which rank-based mobility measures should adhere to and postulated that measures can adhere to one or multiple of these axioms, but never to all of them (these axioms will be addressed in more detail in the subsequent section). Shorrocks then discussed several mobility measures and showed how they fared in satisfying the axiomatic properties. This opened up the floor for the empirical use of rank-based mobility measures. Indeed, today these measures have been used to track the relative mobility in various economies~\cite{JanttiJenkins2015}. Nevertheless, the rank approach is not without limitations. Specifically, rank-based measures indicate mobility whenever there are changes in the positions of at least two individuals. As a consequence, they are usually inconclusive about the source and direction of mobility. To illustrate this, consider the change in rank between two individuals. Rank-based measures will indicate mobility in this case, but we are not sure whether the mobility is because the wealth of one of the individuals increased, or the wealth of the other individual decreased. It could also be the case that the wealth of both individuals increased or decreased, but the rate of change may be higher for a certain individual.

The absolute approach investigates mobility through the direct changes in the wealth of an individual between two time points. A standardly used measure for this kind of mobility is the earnings elasticity, defined as the slope of the regression between the wealth in an initial year and the wealth in a final year~\cite{aaronson2008intergenerational,blanden2013cross,corak2013income}. These measures usually do not have the limitation of the rank-based measures as they will indicate mobility whenever there is a change in the magnitude of wealth of at least one individual. However, they cannot tell us much about the changes in the social ordering. Therefore, studies have usually coupled results derived from these measures with rank-based measurements~\cite{JanttiJenkins2015}.

A common feature of both approaches is that they look at wealth of each individual within a population at an initial date and at a final date. They calculate the change in individual wealth either in raw or rank values, and then aggregate these values into a measure of economy-wide mobility. This aggregation can be seen as a limitation of both approaches as it may lead to results which underestimate or overestimate the mobility of certain groups of individuals. For example, studies have shown that mobility is dependent on gender~\cite{ferreira2012economic}, ethnicity~\cite{bloome2011cohort}, and many other socio-economic factors~\cite{chetty2014land}. These factors are usually not accounted for in both rank-based and raw-based measurements. To reduce this limitation, the more recent literature has developed measures that 
disaggregate mobility to a higher resolution. For instance, Bhattacharya and Mazumder~\cite{bhattacharya2011nonparametric} suggest the use of a measure that disaggregates changes to the level of a demographic group. Yet, still these measures do not tell us the feasibility of everyone in the group to change their wealth.

As we will show subsequently, this is not the case with measures of mixing. These measures do not aggregate the individual values but instead construct indicators that describe the ability of each individual member of the population to move across the wealth distribution. The idea of mixing is frequently used in probability theory and statistical physics for tracking the convergence of probability distributions to their stationary values~\cite{levin2017markov,mcdonald2011estimating}. Recently, it has also been applied in the economic inequality literature to evaluate the ability of mathematical models of income and wealth to mimic the speed of convergence towards the tail of the real world distribution~\cite{gabaix2016dynamics}. 

Our contribution complements all these studies on economic mobility and physical measurements. Its distinguishing result is showing how mixing can be used in tandem with measures of mobility for drawing conclusions about the extent of mobility across the whole distribution.

\section{Mixing and mobility}
\label{sec:mixing-time}

\subsection{Definition}

Mixing describes the property of a dynamical system being strongly intertwined. In physical terms, this means that, for any set of particles in a dynamical system, the fraction of the particles found within a particular region in the \textit{phase space} (the space of the variable $x$ characterizing the particles, wealth in our case), is proportional to the volume of that region in the phase space. Figuratively, we can think of an economy as a cup of coffee and of some person's wealth as milk poured in the coffee. If the system is mixing, then the milk will spread across the coffee over time and eventually it will be spread equally in the cup.
 
To define and analyze mixing, we will use the individual as our basic unit of observation. Nevertheless, we point out that other units such as households can be used to measure mixing (In fact we will use the household as our unit of observation in the empirical analysis). On this basis, we define mathematically mixing as follows. Let $x_i(t)$ denote the wealth in year $t$ of the $i$-th individual in a population of size $N$. Moreover, let $y_i = x_i / \langle x \rangle_N$, where $\langle x \rangle_N = \sum_i x_i / N$  is per capita wealth, be the \textit{rescaled} wealth of individual $i$ and $P(y(t),t)$ be the probability density function that describes the distribution of rescaled wealth in the same year, with the initial condition being a Dirac delta function with a mass centered at 1, i.e., $P(y(0),t_0) = DiracDelta(y-1)$. If the economy is mixing, then starting from an initial year $t = 0$ in which everyone has \textit{identical} rescaled wealth $y(t_0) = 1$, then the distribution $P(y(t),t)$ will in each subsequent time point resemble more and more a predefined target wealth distribution $P^*(y)$. More importantly, it will eventually converge to the target distribution. A standard way for evaluating this property is through the $\beta$-mixing coefficient. The coefficient measures the total variational distance between the wealth distribution in year $t$ and the target distribution, i.e.,
\begin{align}
    \beta(t) &= || P(y(t),t) - P^*(y) ||,
    \label{eq:var-distance}
\end{align}
where $||g(y)|| = \int |g(y)| dy$ is the $L^1$ norm of $g$. Formally, the wealth dynamics is said to be ``$\beta$-mixing'' if $\beta^* = \lim_{t \to\infty} \beta(t) = 0$ \cite{drees2000weighted,stojkoski2021autocorrelation}. 

In economic terms, mixing implies that the system does not discriminate between individuals on the basis of their history: it is possible for everyone to change their position in the distribution over time. 

Although mixing is naturally linked to economic mobility, as defined by standard measures of mobility, the two are not always the same: Mobility, fundamentally represents the ability of individuals to change their wealth or rank within the wealth distribution. As we will see in~\Sref{measures}, there is always a relationship between measures of mixing and standard mobility measures, when mixing exists in the system. However, the standard measures may still indicate that there is some level of mobility even when mixing does not occur. For instance, a finite value of $\beta(t)$ in a non-mixing economy can still signify the presence of mobility, reflecting the extent to which individuals are moving within the distribution. Yet, the existence of such mobility, a degree of which always exists practically, does not guarantee mixing, as this observation depends on the existence of mobility between \textit{every} quantile in the wealth distribution~\cite{Mcfarland1970}.

\subsection{Examples for mixing measures}
\label{sec:relaxation-time}
Any statistical measure that is derived from the properties of $\beta(t)$ can be interpreted as a measure for the degree of mixing within an economy. In this work, we consider two such measures: 1) the relaxation time, and 2) the mixing time. 

By assuming an exponential growth model (which is a common assumption in wealth dynamics), we define the relaxation time $T_{rel}$ as the reciprocal of the rate of convergence, $r$, towards the target distribution, i.e.,
\begin{align}
T_{rel} &= 1/r,    
\label{eq:relaxation-time}
\end{align}
where 
\begin{align}
r &= - \lim_{t \to \infty} \frac{1}{t} \log \beta(t).
\label{eq:convergence-rate}
\end{align}

The mixing time, in contrast, is twice the time point at which $P(y(t),t)$ is for the first time statistically identical to $P^*(y)$. Usually, this is defined as the first time point at which the total variational distance is less than a threshold $\varepsilon$~\cite{levin2017markov}, i.e.,
\begin{align}
    T_{mix} &= 2 \times \inf \{t: \beta(t) < \varepsilon\},
    \label{eq:mixing-time}
\end{align}
where $\inf$ is the infimum of the set of all time points $t$ at which the total variational distance $\beta(t)$ is less than the threshold $\varepsilon$.
For simplicity, here we will use $\varepsilon = 1/4$ as a threshold. This exact value is most often used in Markov Chains to track the convergence to the steady state~\cite{aldous2002reversible,hsu2015mixing,wolfer2019estimating,wolfer2020mixing}. Additionally, this choice leads to approximate equivalence between the Mixing time and the Relaxation time in the theoretical model that we study (see~\ref{sec:rgbm-numerical-mixing-time}). We nevertheless stress out that other thresholds, which may lead either to more restrictive or loose definitions for mixing, can also be used in practice. 

Both measures are naturally measured in years and provide a characteristic timescale over which individuals mix into the wealth distribution. In the coffee analogy, they would quantify the time required for the milk to blend with the coffee. This enables the measures to be used for appropriate comparison between different time periods and economies. The relationship between these two measures and mobility is intuitive. That is, when the relaxation time or the mixing time is short relative to a relevant window of observation, then there is high wealth mobility and strong mixing. Slow relaxation and mixing times are interpreted as indicating low mobility and weak mixing. More importantly, when the relaxation time and mixing time are infinite, then wealth is not a mixing observable, but this does not mean that there is no mobility. 

Both measures, however, exhibit different features when it comes to their estimation procedures. On the one hand, the relaxation time can be easily estimated in theoretical models by exploiting their spectral properties. This measure, however, is difficult to quantify in practice. This is because when the relaxation time is estimated using empirical data, one uses an estimate for the wealth distribution based on a histogram~\cite{mcdonald2015estimating}. Histograms can resemble the theoretical distribution only to a certain extent since they are a finite sample size approximation. As a result, $\beta(t)$ will not always decay to zero, and it will be difficult to numerically track the rate of convergence.

On the other hand, the mixing time  can be easily estimated using empirical data because it does not rely on the model assumptions. Yet, its analytical estimation in theoretical models can be an exhausting task as it involves advanced mathematical apparatus~\cite{wolfer2019estimating}.

Therefore, in the theoretical analysis described in~\Sref{rgbm} we will use the relaxation time as a measure of mixing, whereas in~\Sref{empirical-results} we will use the mixing time to devise our empirical method for quantifying the degree of mixing in a real economy. We refer to~\ref{sec:rgbm-numerical-mixing-time} for more background about the relationship between mixing time and relaxation time.

\subsection{Axiomatic properties of mixing measures}

As a means to understand the differences in the information provided by measures of mixing and measures of mobility, we compare the axiomatic properties of the relaxation time and the mixing time to those of standard measures of mobility. A mobility measure $M\big( \left\{ \mathrm{f}\left(Y(t)\right) \right\}_{t \in \left[t_0, t_f \right]} \big)$ is a continuous real function that maps monotonic transformations $\mathrm{f}(\cdot)$ of observations of the population wealth $Y(t) = \left[ y_{i}(t) \right]$, observed between periods $t_0$ and $t_f$, into a single value. Specifically, in the case of the relaxation time and the mixing time, \(\mathrm{f}\left(Y(t)\right)= y(t)\). Moreover, given the practical limitation that \(t \to \infty\) is not feasible in real data (but we have it in the definition of mixing), we suppose that \(t_f\) be significantly larger than \(t_0\) and use approximate definitions for $M$ function describing the relaxation and mixing time. Namely, by keeping in mind that we are working with finite data, and by combining equations~\eqref{eq:var-distance},~\eqref{eq:relaxation-time}, and~\eqref{eq:convergence-rate}, we derive the expression for the mobility measure in terms of the relaxation time as 
\[M\big( \left\{ Y(t)\right\}_{t \in \left[t_0, t_f \right]} \big) = -\frac{t_f}{\log ||P(Y(t_f))-P^*(y)||}.\]
Similarly, utilizing equations~\eqref{eq:var-distance} and~\eqref{eq:mixing-time}, the expression for the mobility measure in terms of the mixing time is given by
\[M\big( \left\{ Y(t)\right\}_{t \in \left[t_0, t_f \right]} \big) = \begin{cases} 
2 \times \min \{t \in\left[t_0, t_f \right]: ||P(Y(t_f))-P^*(y)|| < \varepsilon\} & \text{if such } t \text{ exists}, \\
\infty & \text{otherwise}.
\end{cases}\]

For the comparison we utilize three such measures: the Spearman's rank correlation, the intragenerational earnings elasticity (IGE), and the Gumbel copula~\cite{berman2022absolute}. In all of the measures that we study, lower values indicate higher mobility. More details about the technical background for these mobility measures is given in~\ref{sec:standard-mobility-measures}. 

We consider four axiomatic properties that a mobility measure may have \cite{Shorrocks1978,cowell2018measuring}: 1) normalization, 2) monotonicity, 3) period dependence, and 4) distribution dependence. In Table~\ref{tab:mobility-properties}, we detail the properties that are satisfied by the standard mobility measures which we study. The table also gives the properties of the relaxation and mixing times, when these measures are used to evaluate the mobility within an economy.

\begin{table}[ht]
\caption{\textbf{Properties of mobility measures and relaxation time.\label{tab:mobility-properties}}}
\resizebox{\textwidth}{!}{%
\begin{tabular}{l||ccccc}
\textbf{Measure }  & \multicolumn{4}{c}{\textbf{Property}} \\\hline
           & Normalization & Monotonicity & P. dependence & D. dependence \\\hline
Relaxation time         & $\times$             & $\times$            & $\times$                 & \checkmark           \\
Mixing time         & $\times$             & $\times$            & $\times$                 & \checkmark           \\
Spearman Correlation & \checkmark             & \checkmark            & \checkmark                 &  $\times$                         \\
IGE                  & $\times$             & \checkmark            & \checkmark                 & \checkmark                           \\
Gumbel parameter     & \checkmark             & \checkmark            & \checkmark                 & $\times$   \\                
\end{tabular}}
\end{table}

\paragraph{Normalization property:}

The normalization property imposes a closed interval bound on the mobility measure. This is, by definition, not satisfied by our measures of mixing as they are measured in years. Relaxation and mixing times could take 5 years, 100 years, or even an infinite amount of time. However, we state that normalization is just a standard procedure that can be implemented on any measure, and the relaxation and mixing times can be easily adjusted to satisfy this property. For example, both measures can be transformed to an index $w$ valued between 0 and 1, where 0 indicates the highest mobility, by defining $w = e^{-1/T}$. A similar normalization was introduced by Shorrocks \cite{Shorrocks1978} for measures derived from transition matrices. We purposely refrain from the transformation procedure because we want to emphasize whether some value of relaxation time is high or low in terms of the years required for an individual to move across the whole distribution. 

\paragraph{Monotonicity property:}

The monotonicity property indicates that whenever a new structure is imposed on the wealth dynamics of some individuals affecting the wealth observations in any time point except the initial observation $t_0$, then this is reflected in the value of the mobility measure. We can formalize this property by defining the new structure as $Y'(t)$ and imposing a strict ordering $$\left\{ Y'(t) \right\}_{t \in \left[t_0, t_f \right]} \succ \left\{ Y(t) \right\}_{t \in \left[t_0, t_f \right]}$$ whenever $$||\mathrm{f}(y'(t)) - \mathrm{f}(y(t_0))||_2 > ||\mathrm{f}(y(t)) - \mathrm{f}(y(t_0))||_2,$$ with $||\cdot||_2$ being the $L^2$ norm, for at least one time point $t$ that is used for estimation of the mobility measure. Monotonicity implies that if $$\left\{ Y'(t) \right\}_{t \in \left[t_0, t_f \right]} \succ \left\{ Y(t) \right\}_{t \in \left[t_0, t_f \right]},$$ then $$M\big( \left\{ \mathrm{f}\left(Y'(t)\right) \right\}_{t \in \left[t_0, t_f \right]} \big) < M\big( \left\{ \mathrm{f}\left(Y(t)\right) \right\}_{t \in \left[t_0, t_f \right]} \big).$$ 

Every measure of mixing will not satisfy this property. The inability to satisfy the monotonicity property arises because monotonicity implies that, if there is mobility between certain quantiles of the wealth distribution, it will be translated as existence of mobility in the standard measures. The three described measures represent aggregated values of the changes in the wealth of the individuals which constitute the population between two time periods. Measures for mixing, on the other hand, are not monotonic unless there is already mobility between every quantile. In every other case they will imply zero mobility.

To show this, assume a non-mixing system in which the population distribution converges to a distribution $P_{\kappa}(y)$ that is parametrized by a single parameter $\kappa$. Furthermore, let $\beta^*_{\kappa} = || P_{\kappa}(y) - P^*(y)||$ be a monotonically decreasing function of $\kappa$ and let $||P_{\kappa}(y) - P^*(y)|| > 1/4$ . Then, $Y(t) \sim P_{\kappa}(y)$ for large enough $t$, and by construction the relaxation and mixing times will be infinite. Now, consider the case where we slightly perturb the population distribution to $P_{\kappa + \epsilon}(y)$, where $\epsilon > 0$, but we still let $\beta^*_{\kappa+\epsilon} > 1/4$. Then, $Y'(t) \sim P_{\kappa + \epsilon}(y)$ and $||\mathrm{f}(y'(t)) - \mathrm{f}(y(t_0))||_2 > ||\mathrm{f}(y(t)) - \mathrm{f}(y(t_0))||_2$ for large enough $t$. Nevertheless, we will still have that $\beta^*_{\kappa+\epsilon} > 1/4$, implying that the mixing and relaxation times are also still infinite ($M\big( \left\{ \mathrm{f}\left(Y'(t)\right) \right\}_{t \in \left[t_0, t_f \right]} \big) = M\big( \left\{ \mathrm{f}\left(Y(t)\right) \right\}_{t \in \left[t_0, t_f \right]} \big)$).

\paragraph{Period dependence property:} The period dependence property implies that the ratio of the same mobility measure estimated at two different end points $t_f$ and $t_f' = t_f + \Delta$, is a function of the temporal difference $\Delta$. This can be formalized as $$\frac{M\big( \left\{ \mathrm{f}\left(Y(t)\right) \right\}_{t \in \left[t_0, t_f \right]} \big)}{M\big( \left\{ \mathrm{f}\left(Y(t)\right) \right\}_{t \in \left[t_0, t_f+\Delta \right]}\big)} = C(\Delta),$$ where $C(\cdot)$ is a monotonically increasing function. 

This property tells us that if the economic conditions are not changing, then for larger observation periods we should also observe higher mobility. Period dependence, however, is not  satisfied by our measures of mixing by definition. The relaxation and mixing times will suggest to identical degree of mobility independent of whether our observation period is 10, 20, or 60 years. This is because measures of mixing look at the limiting case of $\beta(t)$. Hence, they are either based on large time-series observations ($t_f \to \infty$) or large $t$ observations are approximated via a regression fit (see \Sref{empirical-results}). Conversely, the Spearman rank correlation, the IGE, and the Gumbel copula quantify the mobility in ``per $\Delta$ years'' and therefore their magnitude will always be dependent on $\Delta$.

\paragraph{Distribution dependence property:}
Finally, distribution dependence implies that the mobility predicted by the measure is a function of the wealth distributions $P(y(t),t,t)$ that are used for its estimation, i.e.,
$$M\big( \left\{ \mathrm{f}\left(Y(t)\right) \right\}_{t \in \left[t_0, t_f \right]} \sim \Phi\big(P(y(t),t)\big),$$
where $\Phi\big(\cdot\big)$ is a continuous function.

Obviously, this property is captured by measures of mixing as they are defined through the evolutionary characteristics of the wealth distribution: In measures of mixing $\Phi\big(P(y(t),t)\big) = \beta(t)$. This dependence indicates that measures of mixing will not be interpreted similarly when the underlying wealth distribution remains unchanged, and when it becomes more and more unequal. Interestingly, from the three standard measures, only the IGE is dependent on the shape of the wealth distributions that are used for its estimation, whereas the other two measures are not.

\section{Mixing in a simple model of an economy}\label{sec:rgbm}
\subsection{Reallocating geometric Brownian motion}

To illustrate the application of mixing in economic systems, we use reallocating geometric Brownian motion (RGBM), a simple model for wealth dynamics~\cite{BermanPetersAdamou2019}. Under RGBM, wealth is assumed to grow multiplicatively and randomly, in addition to a simple reallocation mechanism. The dynamics of the wealth of person $i$ are specified as
\be
\mathrm{d} x_i = x_i \left( \mu \mathrm{d}t + \sigma \mathrm{d}W_i \right) - \tau \left( x_i - \langle x \rangle_N \right) \mathrm{d}t\,,
\label{eq:rgbm}
\ee
with $\mu > 0$ being the drift term, $\sigma > 0$ the fluctuations amplitude, and $\mathrm{d}W_i$ is an independent Wiener increment, $W_i(t) =\int_0^t \mathrm{d}W_i$. The parameter $\tau$ quantifies the rate of reallocation of wealth. It implies that, in every time period $\mathrm{d}t$, everyone in the economy contributes a fraction $\tau\mathrm{d}t$ of their wealth to a central pool. The pool is then shared equally across the population. This parameter encapsulates multiple effects, \eg collective investment in infrastructure, education, social programs, taxation, rents paid, or private profits.

Under RGBM, the average wealth in a large population grows like $e^{\mu t}$. Rescaling by $e^{\mu t}$, the dynamic behavior of RGBM is strictly dependent on the relation between $\tau$ and $\sigma$, and the rescaled wealth can be both mixing and non-mixing. When $\tau > 0$, rescaled wealth in RGBM is mixing, ergodic and stationary. The model exhibits mean-reversion as each $x_i$ reverts to the population average $\langle x \rangle_N$. The dynamics of the rescaled wealth $y_i = x_i / \langle x \rangle_N$ can be described as
\be
    \mathrm{d} y =   y \sigma  \mathrm{d} W - \tau (y - 1)  \mathrm{d}t.
    \label{eq:rescaled-rgbm}
\ee
The stationary (target) distribution of the model is
\be
P^*(y) = \frac{(\zeta - 1)^{\beta}}{\Gamma(\zeta)} \exp{\big(-\frac{\zeta - 1}{y}\big)} y^{-(1+\zeta)}\,,
\label{eq:rgbm-stationary-distribution}
\ee
where $\zeta = 1 + \frac{2 \tau}{\sigma^2}$ and $\Gamma(\cdot)$ is the Gamma function (see Ref.~\cite{BermanPetersAdamou2019}). The distribution has a power-law tail. The exponent of the power law, $\zeta$, is called the Pareto tail parameter, and can be used as a measure of economic equality \cite{Cowell2011}. More importantly, important stylized facts are recovered: the larger $\sigma$ (more randomness in the dynamics) and the smaller $\tau$ (less reallocation), the smaller the tail index and the fatter the tail of the distribution, leading to higher inequality. When $\tau \leq 0$, there is no stationary distribution to which rescaled wealth converges.

\subsection{Relaxation time in RGBM}
\label{sec:relax-time-rgbm}
The estimation of relaxation time in a stochastic process is usually done by investigating the Fokker-Planck equation which describes the evolution of the probability density function. The Fokker Planck equation for the probability density function $P(y,t)$ for the differential equation described in Eq.~\eqref{eq:rescaled-rgbm} is
\begin{align}
       \frac{\partial}{\partial t}P(y,t)= \tau \frac{\partial}{\partial y} \left[(1-y) P(y,t)\right] + \frac{\sigma^2}{2}\frac{\partial^2}{\partial y^2} \left[ y^2 P(y,t) \right].
\end{align}
More details about the properties of RGBM's Fokker-Planck equation can be found in~\cite{BouchaudMezard2000,LiuSerota2017,BermanPetersAdamou2019,stojkoski2021ergodicity}.\footnote{In some of the literature, a different notation is used for the parameters of RGBM and thus the Fokker-Planck equation is slightly different. For example, for the models studied in~\cite{BouchaudMezard2000,LiuSerota2017} one needs to substitute the parameters $J \to \tau$ and $\sigma \to \sigma / \sqrt{2}$ in order to obtain our model.}

The equation is an appropriate generalization of a transition matrix for a discrete space process to processes with continuum of states. Hence, the analysis follows exactly in the same manner: the critical statistic of the Fokker-Planck equation that governs the convergence of $P(y(t),t)$ to $P^*(y)$ is the largest nontrivial eigenvalue of the corresponding Fokker-Planck operator, that is, the second largest eigenvalue.\footnote{It is important to note that the Fokker-Planck equation is synonymous with the Kolmogorov Forward equation in the context of describing the time evolution of a probability distribution for stochastic processes. Additionally, we have employed a change of variables where \( y = \frac{x}{\langle x \rangle} \), as stated earlier in this section.} 
 In a mixing system the largest eigenvalue will be zero, whereas the second eigenvalue, $\lambda_2$, will be \textit{negative}, and $\beta(t) \propto e^{\lambda_2 t}$. We refer to Gabaix et al.~\cite{gabaix2016dynamics} for a detailed mathematical background on estimating the relaxation time in systems where the governing dynamical equation for the stochastic process is known.

An extensive study for the eigenvalues of RGBM was done by~\cite{LiuSerota2017}. The authors showed that the second largest eigenvalue is simply $-\tau$ and, therefore, that the relaxation time will exist only when $\tau > 0$ and it will be equal to $1/\tau$.\footnote{As demonstrated in~\cite{LiuSerota2017}, the relaxation to the stationary distribution can indeed be influenced by the initial conditions. However, in the long run, the relaxation time emerges as a fundamental property of the system, being solely dependent on the second largest eigenvalue, and thus remains invariant to the choice of initial conditions.} The interpretation behind this result is fairly intuitive -- in an economy in which reallocation from the rich to the poor is stronger, mixing is faster. On the other hand, as the reallocation rate approaches zero, relaxation times get longer, and mobility, as defined by the concept of mixing lower. As the model becomes non-stationary for reallocation rates that are equal or less then $0$, the possibility for mixing in the economy disappears. Then, the relaxation time is infinite. We point out that while $\tau <0$ implies no mixing, that does not mean that the standard mobility measures will also indicate no mobility. In fact, for any positive fluctuation amplitude $\sigma$, there will be randomness in the system. This randomness may drive changes in the observed wealth, and as a result the standard measures may display a certain degree of mobility.

\subsection{Relaxation time and standard measures of mobility in RGBM}\label{sec:measures}

In RGBM, the standard measures of mobility depend on both $\tau$ and the fluctuation amplitude $\sigma$, unlike the relaxation time. This is a consequence of the randomness playing a significant role in the wealth dynamics when we consider timescales that are shorter than the time required for the system to relax. In what follows, we describe the relationship between relaxation time and the standard measures of mobility in RGBM.

\paragraph{Spearman's rank correlation:} Spearman's rank correlation is directly related to the relaxation time in RGBM: higher Spearman's correlation implies lower relaxation times. The rank correlation is also dependent on the fluctuation amplitude $\sigma$ and the temporal difference $\delta$ between the two periods that are being compared. Larger values for both parameters lead to greater economic mobility. This can be seen in~\fref{rgbm-standard-measures}A, where we plot the log of the rank correlation divided by $\delta$ as a function of $\tau$ for various noise amplitudes.

\begin{figure}[!htb]
\centering
\includegraphics[width=1.0\textwidth]{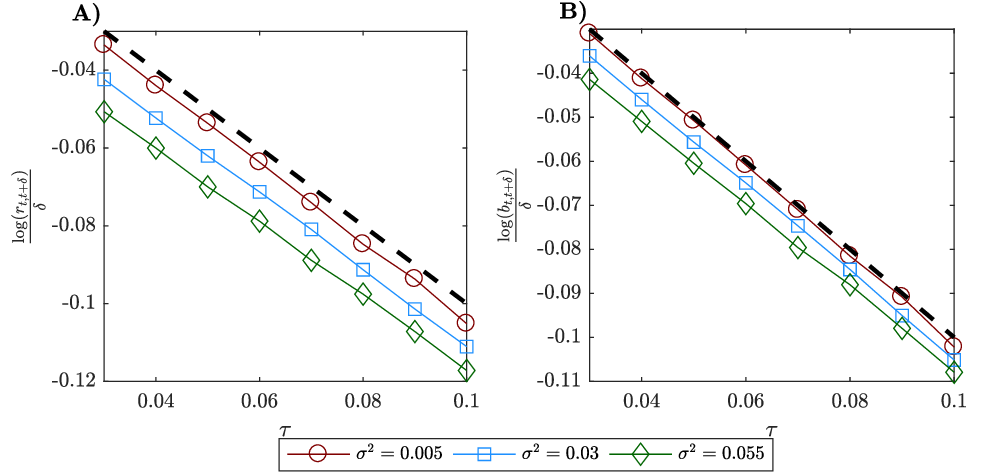}
\caption{\textbf{Relaxation time and standard measures of economic mobility.} \textbf{A)} Log of Spearman's rank correlation divided by the temporal difference as a function of $\tau$. \textbf{B)} Same as \textbf{A)}, only on the y axis is the log of the IGE divided by the temporal difference. \textbf{A-B} The dashed black line has a slope 1. The simulations used $\delta = 20$ years and $N = 10^4$ people.
\label{fig:rgbm-standard-measures}}
\end{figure}

\paragraph{Intragenerational earnings elasticity:} Similarly to the properties of the rank correlation, and as depicted in~\fref{rgbm-standard-measures}B, the IGE depends on both the fluctuation amplitude and the reallocation rate.

\paragraph{Transition matrices:} As evidenced in~\fref{rgbm-wealth-matrices}A, the transition matrices in RGBM reproduce the asymmetric property of the real world transition matrices and are well-approximated by the Gumbel copula (\fref{rgbm-wealth-matrices}B). In \fref{rgbm-wealth-matrices}C we visualize the relationship between Gumbel parameter $\theta$ and the reallocation parameter $\tau$ for various fluctuation amplitudes. We find that there is an inverse relationship between $\theta$ and $\tau$, and the Gumbel parameter slope is further determined by the magnitude of $\sigma$. As $\tau$ increases (relaxation time decreases), the value of the $\theta$ decreases, though disproportionately. 

\begin{figure}[!htb]
\centering
\includegraphics[width=1.0\textwidth]{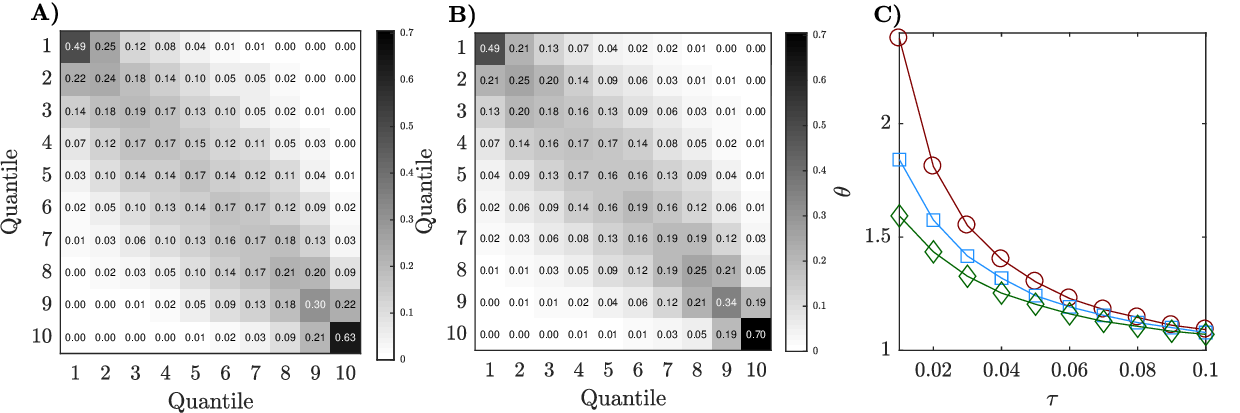}
\caption{\textbf{Relaxation time and wealth transition matrices.} \textbf{A)} Transition matrix for the stationary regime of RGBM estimated with $\tau = 0.02$ per year, $\sigma^2 = 0.01$ per year and $N = 10^4$ people. \textbf{B)} An example for a transition matrix from data simulated from a Gumbel copula whose parameter $\theta$ is chosen to be in accordance with the RGBM parameters used in \textbf{A)}. \textbf{C)} The relationship between the Gumbel copula parameter $\theta$ and the reallocation parameter $\tau$ in the stable state of RGBM. This relationship was estimated by simulating a transition matrix in which $\delta = 20$ years and $N = 10^4$ people.}
\label{fig:rgbm-wealth-matrices}
\end{figure}
\FloatBarrier

\subsection{Mobility measures in non-mixing regimes}

\begin{figure}[!htb]
\centering
\includegraphics[width=1.0\textwidth]{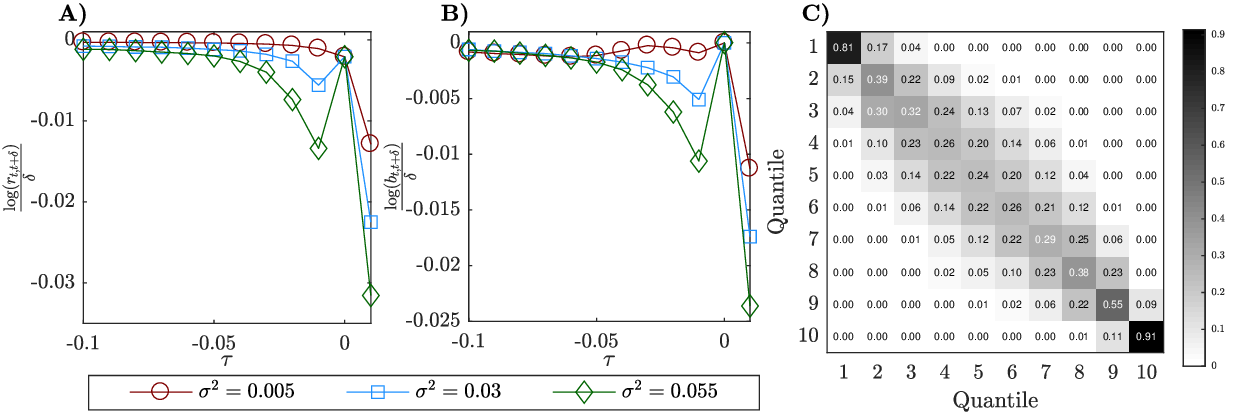}
\caption{\textbf{Mobility in the non-mixing regime of RGBM.} \textbf{A)} Log of Spearman's rank correlation divided by the temporal difference as a function of $\tau$. \textbf{B)} Same as \textbf{A)}, only on the y axis is the log of the IGE divided by the temporal difference. \textbf{C)} Transition matrix in RGBM estimated with $\tau = -0.02$ per year, $\sigma^2 = 0.01$ per year and $N = 10^4$ people. \textbf{A-C} The simulations used $\delta = 20$ years and $N = 10^4$ people. The results are averaged across 1000 model realizations.
\label{fig:rgbm-negative-regime}}
\end{figure}

So far we discussed the relationship between mobility measures and mixing in the stationary regime of RGBM. Studying this relationship in the non-stationary regime, when $\tau < 0$ is impossible as then the system is not mixing~\cite{stojkoski2021ergodicity}. Nonetheless, Spearman's rank correlation, the earnings elasticity and the transition will indicate that there is still some mobility (they will have a value below 1). We visualize this phenomenon numerically in Fig.~\ref{fig:rgbm-negative-regime}. The results for Spearman's correlation and the IGE suggest that in the negative $\tau$ regime, the magnitude of the reallocation rate has a minor effect the extent of mobility. Instead, mobility is mostly dependent on $\sigma$, with larger noise amplitudes implying more mobility. Hence, it can be argued that the existence of mobility in the negative regime is solely a result of the randomness present in the system, but not of the reallocation mechanism. More importantly, because of this, individuals are not able to move across each possible rank. The thorough study of mobility in RGBM in this regime is outside of the scope of this paper and left for future work.

\section{Mixing in real economic systems}
\label{sec:empirical-results}

\subsection{Empirical method for estimating mixing}
There are different ways to estimate mixing in real economic systems. The simplest approach is to assume that the wealth dynamics obeys some stochastic process (e.g., RGBM) and estimate the model parameters using time series data. The estimated parameters can be used to quantify the spectral properties of the process and thus approximate the relaxation time. For example, Berman et al.~\cite{BermanPetersAdamou2019} use the RGBM model and historical data on the share of wealth owned by the top 1\% to estimate the reallocation $\tau$ parameter in the USA economy in the period between 1913 and 2014. The authors find that the reallocation parameter is either negative or a value below $0.02$, implying that either mixing does not occur or that it takes more than 50 years for individuals to experience~\cite{BermanPetersAdamou2019}. This approach, however, is limited in its applicability as it requires strong assumptions about the model governing the dynamics of wealth. For example, in RGBM wealth is redistributed with an equal rate among the individuals, differently from how usually real economic mechanisms work. 

To circumvent the limitations that might arise because of model assumptions, we present a non-parametric empirical method for evaluating the mixing properties of wealth dynamics in an economy and showcase its application using data for the USA. This empirical method is inherently flexible and not model dependent.

To quantify the degree of mixing empirically, we will use the mixing time $T_{mix}$ as our measure of wealth mixing. Additionally the method requires longitudinal data and tracks the time until the distribution of a subsample of the population representing the typical individuals resembles the wealth distribution of the whole economy. 

The method for quantifying the Mixing time is constituted of the following steps.

\paragraph{Step 1. Define targets:} In the first step, we define a relevant target distribution to which the rescaled transformation of wealth converges. This can be either a theoretical distribution (such as the stationary distribution of RGBM) or an empirical estimate for a distribution taken from the histogram of wealth in the country given at a fixed future data.

\paragraph{Step 2. Select the typical individuals:} We then select a subsample consisting of the ``typical'' individuals in the economy. These are the individuals that are closest to the mean rescaled wealth. For simplicity, we formally define this group of individuals as those that have rescaled wealth between $0.95$ and $1.05$. This is similar to the condition of a Dirac delta initial wealth distribution distribution. We point out that other intervals for the wealth of the typical individuals can be used, as long as the subsample distribution resembles a Dirac delta function.

\paragraph{Step 3. Track the wealth dynamics:} Next, we track the subsample wealth over time and quantify the difference between the subsample wealth distribution and the target distribution with the total variational distance. In general, the total variational distance will exhibit two states. First, there will be a transitory time state during which the subsample distribution will converge to a stationary distribution. In this regime $\beta(t)$ will decay. After this transitory phase, there will be a stable state. In the stable state, the log of the distance reaches a plateau $\beta^*$ and fluctuates around this plateau. If the economy is non-mixing, then the total variational distance will converge to a value larger than $1/4$. In contrast, if the economy is mixing (with respect to the target distribution), then $\beta(t)$ will converge to a value much lower than $1/4$. The first time point at which $\beta(t)$ will be less than $1/4$ will be, in general, smaller than the relaxation time, and if this time point is multiplied by two we will recover a value that is close to the relaxation time (see~\ref{sec:rgbm-numerical-mixing-time}).

\Fref{mixing-time} summarizes the method in a fictive example of a mixing economy. The blue line is the log of the total variational distance as a function of time. The dashed black line is the slope of the relationship between the statistic and time during the two different states. The inset plots provide snapshots for the wealth distribution of the subsample (i.e., estimates based on a histogram) at different time points (red dashed lines). For comparison, the snapshots also include the form of the selected target distribution (black line). Notice that initially, at $t_1$, the subsample distribution is very narrow and does not resemble the target distribution. The wealths in the subsample evolve, and in $t_2$ and $t_3$ the subsample distribution becomes closer to the target distribution. Eventually, the subsample distribution resembles the target distribution. 

\begin{figure}[!htb]
\centering
\includegraphics[width=1.0\textwidth]{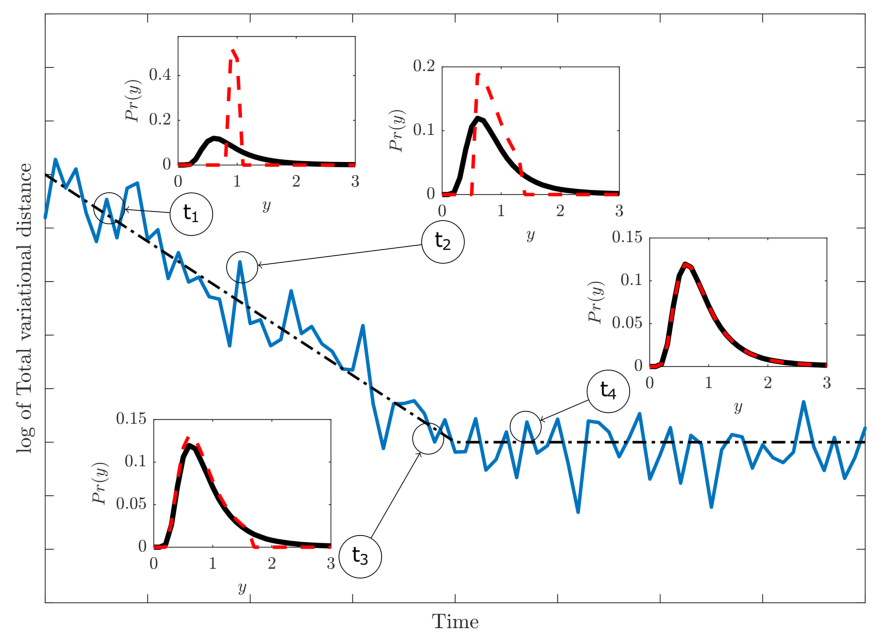}
\caption{\textbf{Quantifying mixing in an empirical system.} The blue line shows the log of the total variational distance between the subsample and the target wealth distribution. The black dashed line is the slope of the line which describes the relationship between the log of the distance and time, estimated separately for the decay period and the stable state period. The inset plots give snapshots for the empirical form of subsample distribution (red dashed line) and the target distribution at different points in time $t_1 < t_2 < t_3 < t_4$. 
\label{fig:mixing-time}}
\end{figure}
\FloatBarrier

\subsection{Mixing in the United States}

We demonstrate the application of our method using wealth data taken from the Panel Study of Income Dynamics (PSID) dataset. The PSID is a comprehensive longitudinal survey of over 18,000 individuals living in 6,000 families in the United States. It measures economic, social, and health factors over the life course of the families. Since 1984 it also collects data on the wealth of the families. Up until 1999 these data were collected quadrennially, and after this year the wealth data are collected biannually.

Because the PSID is based on survey data it has some limitations that restrict its practical application. In particular, it has a small cross-sectional size: in each year only around 6000 families were interviewed. This issue is slightly resolved by providing population weights to each observation~\cite{biemer2012weighting}. These weights, nevertheless, can lead to measurement errors that result in restricted coverage of the top of the distribution~\cite{bound2001measurement}. Also there is a general difficulty in the evaluation of wealth in surveys, which creates large measurement errors and thus tend to increase estimated mobility. Indeed, all these limitations reduce the cross-sectional variation in the data and may lead to an underestimation of the mixing time. Therefore, the analysis presented here is only pedagogical. We believe that with the rapid development of data gathering methods and the improved understanding of wealth dynamics within a population, some of this shortcoming will be overcome, yielding a more in-depth interpretation of mixing in real economies.

Keeping in mind the limitations, we use the data to estimate the mixing times in the United States economy. To reduce the potential noise in the data, in our analysis the unit of observation is a household. For each survey year up to 2009, we provide estimates for the time needed by the wealth of the typical households to statistically resemble the wealth distribution observed in 2019 (Step 1). This implies that we estimate 9 mixing times where as initial points we use the wealth of the typical households in a survey year. In each case, we set the wealth distribution observed in 2019 as the target distribution to which the subsample wealth should converge. There are two advantages of using this target distribution. First, it is the final year for which we have data. As such, observations have the longest available time to converge to it.  Second, by setting this target distribution for different initial time points, we are able to distinguish changes in the mixing time that arise due to the dynamics of wealth from changes in inequality (shapes of the wealth distribution). 

In order to create as realistic as possible estimates for the mixing time, we further clean the data from noisy observations. In particular, to quantify the wealth distribution in each year we use data only on households that have data in every survey year~\cite{berman2022absolute}. This removes the non-surviving families from the analysis and makes the observed distributions between years comparable. To make the the resulting sample comparable to the population data, we follow Perez et al.~\cite{perez2010re} and use a re-weighting procedure. Moreover, to create the potential subsample, we control for the age of the households use data only on households in which the age of the head was between 30 and 35 years in 1984. This is because empirical observations have shown that age affects the ability of individuals to move across the wealth ladder~\cite{jianakoplos1997wealth,steckel2006wealth}. We choose households where the head is between 30 and 35 years of age because these are households representing relatively young families at the beginning of observation and as such should be more representative of mixing~\cite{sewell2003we,read2014social}. Then, we use these data to create a subsample describing the typical households for each survey year between 1984 and 2009 (households with rescaled wealth between $0.95$ and $1.05$ and with age of head between 30 and 35 years in 1984) (Step 2). We track the evolution of their wealth and in each time point with available data (survey year) we compare it through the total variational distance with the target distribution (Step 3). The data in the PSID sample  More details about the cleaning and preprocessing of the data can be found in ~\ref{sec:psid-data-cleaning}.

We visualize our empirical findings in~\Fref{empirical-mixing-time-1}. In~\Fref{empirical-mixing-time-1}A-C we display the evolution of $\beta(t)$ as a function of time for three different starting years: 1984 (\Fref{empirical-mixing-time-1}A), 1999 (\Fref{empirical-mixing-time-1}B), and 2009 (\Fref{empirical-mixing-time-1}C). In each case, we observe that the total variational distance has not reached the $1/4$ limit. This suggests that the mixing time must be larger than the difference between the target year and the initial year. To provide an estimate for the value of the mixing time, we assume that the decay will continue until it reaches a target $\beta^*$, and fit a non-linear regression of the form $\beta(t) ~ b_0\exp\left[-b_1 t\right] + \beta^*$. We consider three different choices for $\beta^*$. First, we assume that $\beta^*$ is an additional parameter in the regression, and we learn it using the available data. In this case, we may infer that the mixing time does not exist (i.e., if we get an estimate $\beta^* > 1/4$ then the mixing time is infinite). Second, we suppose that $\beta^* = 1/4$. In this case we assume that $\beta^*$ converges to the largest value that allows mixing to exist. This will lead to the lowest decay rate, and thus can be seen as an upper bound for the mixing time under the assumption that wealth in the USA economy is is mixing. Lastly, we assume that $\beta^*$ will converge to 0. In this case, the decay rate will be the largest, and hence the result can serve as an estimate for the lower bound for the mixing time under the assumption that the wealth dynamics are mixing.

\begin{figure}[!htb]
\centering
\includegraphics[width=\linewidth]{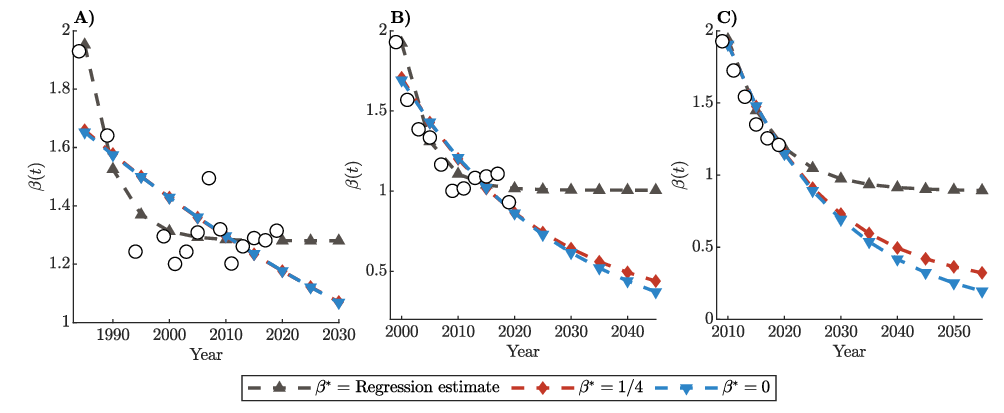}
\caption{\textbf{Total variational distance in the USA estimated using the Panel Study for Income Dynamics.} \textbf{A)} Total variational distance as a function of time in which the target distribution is the USA wealth distribution in 2019, and the subsample data come from the 1984 PSID. \textbf{B)} Same as \textbf{A)}, only the subsample data come from the 1999 PSID. \textbf{C)} \textbf{B)} Same as \textbf{A)}, only the subsample data come from the 2009 PSID. \textbf{A-C} The dashed lines correspond to estimated regression lines for the decay of $\beta(t)$ for different choices of stationary total variational distances (see the legend). 
\label{fig:empirical-mixing-time-1}}
\end{figure}

\Fref{empirical-mixing-time-2} gives the estimated mixing times for each year that we study. We find that if we let the model learn the stationary $\beta^*$, the mixing time is infinite in each year (the black upward-pointing triangle in~\Fref{empirical-mixing-time-2}). When we individually set the stationary $\beta^*$ and assume that the economy is mixing, we get that the mixing time are relatively large when compared to the usual working life of an individual member of a household. The largest mixing times were observed in the 1980s, when the case of $\beta^* = 1/4$ stationary value suggests a mixing time of around 600 years when the subsample is selected in 1984 and around 400 years when the subsample is selected in 1989. When we set $\beta^* = 0$, the mixing time decreases to around 400 years for 1984 and around 300 years for 1989. As we move towards more recent years, the mixing time decreases and after 1999 it appears to be slightly less than 200 when $\beta^* = 1/4$ and around 100 when $\beta^* = 0$. These results suggest that mixing times have been decreasing for the cohort households that had age of head between 30 and 35 years in 1984. Nevertheless, it is still very difficult for these households to move across the whole wealth ladder in the USA.

\begin{figure}[!htb]
\centering
\includegraphics[width=17cm]{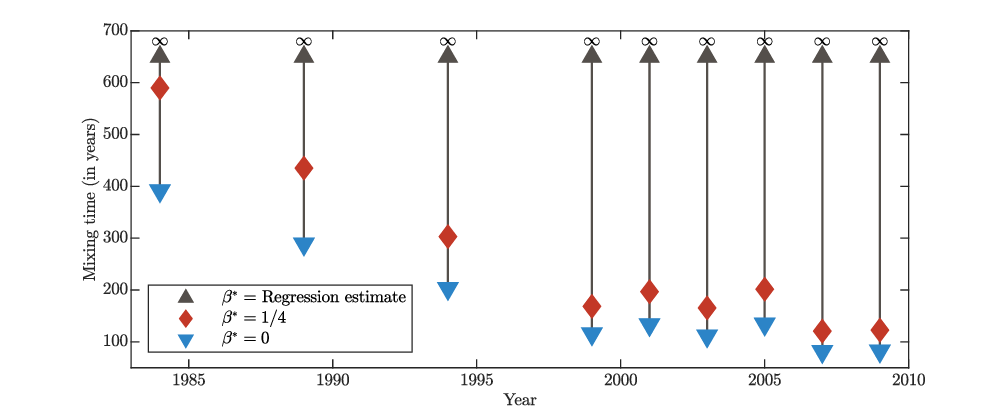}
\caption{\textbf{Mixing times in the USA estimated using the Panel Study for Income Dynamics.} Mixing time estimates for each PSID survey data between 1984 and 2009 for three different stationary total variational distances $\beta^*$. 
\label{fig:empirical-mixing-time-2}}
\end{figure}

\section{Discussion and Conclusion} \label{sec:discussion}

In this paper, we introduced the physical concept of mixing as a relevant phenomenon for quantifying the mobility of the typical individual. We showed that standard mobility measures do not account for this property, whereas measures of mixing do. This is because, every mobility measure that quantifies the degree of mixing will not be monotonic. This particular characteristic of certain measures allows them to quantify whether the system is in a mixing state or not. Whenever the system is not mixing, measures of mixing will suggest no mobility. When the system is mixing, then there is no discrimination between individuals on the basis of their history: it is possible for everyone to move between any ranks in the distribution, and this will happen with certainty in the long run. 

We also showed how mixing can be used empirically by designing a non-parametric approach for estimating the mixing time. By using data stemming from PSID, we were able to provide estimates for the mixing time for US households for the period between 1984 and 2009.

These results can help policy makers in designing interventions that target the well-being of the typical individuals in the economy and can be used in tandem with standard measures of mixing for deriving holistic interpretation about the extent of mobility in the economy. Nevertheless, the mixing approach presented here for measuring the mobility of the typical individual is not without its limitations.

First, mixing is relative with respect to the existence of a transformation of wealth which converges to a target distribution. Here we covered the case of ``rescaled wealth''. But even if we observe that rescaled wealth is non-mixing, another transformation of wealth might exist which is mixing, and measures in terms of it might suggest that there is mixing in the economy. Then mobility across the whole distribution exists, but can be defined in terms of an another concept~\cite{JanttiJenkins2015}. For example, mobility can be defined in terms of growth of wealth or in terms of reduction of the unpredictability of wealth dynamics. Different concepts also require different analytical approaches, as they illuminate the distinct extent to which mobility is socially desirable. In other words, depending on the definition of economic mobility, an increase in economic mobility will not always translate into increased economic welfare. Hence, discovering the relevant wealth transformation and investigating its mixing properties is extremely important for policymakers to produce adequate measures for optimizing the mobility within an economy. 

Second, the empirical method for estimating the mixing properties of an economic system introduced here can be an expensive method to perform in reality as it requires a detailed track for the wealth of a particular set of individuals. Nevertheless, we believe that with the rapid development of data gathering methods and the improved understanding of wealth dynamics within a population, some of this shortcoming will be overcome, yielding a more in-depth interpretation of mixing in real economies.

Lastly, in this paper we discussed mixing in terms of wealth mobility, but did not go into detail about the properties of mixing in income dynamics~\cite{stojkoski2021geometric,stojkoski2021income}. While the estimation procedures for mixing will be the same, the empirical mixing properties of income might be significantly different from those of wealth. Hence, the resulting implications could have a non-trivial impact for economic policies. In this context, studying the empirical relationship between mixing and measures of income mobility may represent a fruitful avenue for future research. 

Despite these limitations, mixing offers a unique view for the measurement of the mobility across the whole wealth distribution. As such, the results presented here advance our understanding on the dynamics of wealth, and should motivate new research on coupling mixing with standard measures of mobility for drawing a more comprehensive picture about economic mobility.



\section*{Declaration of interest}

 The author declares no competing interests.

\clearpage

\appendix

\renewcommand{\thesection}{Appendix \arabic{section}}

\section{Comparison of mixing time and relaxation time in RGBM}\label{sec:rgbm-numerical-mixing-time}

\setcounter{equation}{0}
\setcounter{figure}{0}
\setcounter{table}{0}
\setcounter{subsection}{0}
\setcounter{subsubsection}{0}
\makeatletter
\renewcommand{\theequation}{A1.\arabic{equation}}
\renewcommand{\thesubsection}{A1.\arabic{subsection}}
\renewcommand{\thetable}{A1.\arabic{table}}
\renewcommand{\thefigure}{A1.\arabic{figure}}

A great body of literature has investigated the relationship between mixing time and relaxation time for ergodic Markov Chains, but its properties in continuous time processes and/or space processes remains less explored. 

To provide a glimpse on this relationship in mathematical models of wealth dynamics, we use RGBM and numerically compare the mixing time and relaxation times. We use these results to motivate our definition for the mixing time measure.

We design our numerical study as follows. First, we simulate a population of RGBM using various parameter setups and calculate the total variational distance. Then, we calculate the relaxation time (analytically) and the mixing time (using the method described in \Sref{empirical-results} and a threshold $\varepsilon = 1/4$ ) and compare their values.

Typical trajectory results for different reallocation rates, noise amplitudes, and subsample sizes are shown in~\Fref{rgbm-mixing-time}A-B. It can be easily noticed that the reallocation rate uniquely determines the relaxation time, whereas the noise amplitude has no effect, as argued in~\Sref{rgbm}. Interestingly, the subsample size also determines the convergence to the stable state in the system as it is proportional to the value of the stationary total variational distance $\beta^*$ (\Fref{rgbm-mixing-time}C). Lower subsample sizes lead larger $\beta^*$. This is because the estimation of the distance relies on the differences between the empirical distribution function (i.e., histogram), and the distribution function for the target stationary wealth distribution. Due to the subsample size always being a finite number, in empirical calculations, there will be differences between the empirical distribution and the target distribution, which will be translated in a positive total variational distance. As the subsample size increases, in the limit as the subsample size goes towards infinity, the differences will disappear. Indeed, we observe that this dependence decays exponentially and it is already insignificant for a population size that is around 10\% of the population. This subsample size is comparable to our empirical subsample created in \Sref{empirical-results}, and therefore we do not correct for the subsample size in our definition for mixing time. Nevertheless, in situations with smaller sample we might need adjustments. We leave the investigation for the exact relationship between mixing time and the subsample size as an interesting future exercise.

\begin{figure}[!htb]
\centering
\includegraphics[width=1.0\textwidth]{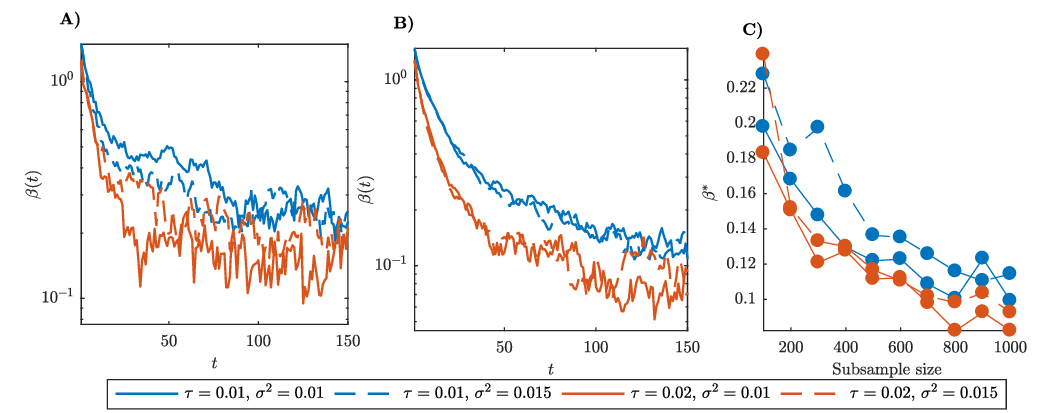}
\caption{\textbf{Total variational distance in RGBM.} \textbf{A)} Total variational distance as a function of time for a realization of an RGBM process with a subsample size of $10^2$ for different noise amplitudes and reallocation parameters. \textbf{B)} Same as \textbf{A)}, only with a subsample size of $10^3$ people. \textbf{C)} The stationary total variational distance as a function of the subsample size for various $\sigma^2$ and $\tau$. In the simulations $N = 10^5$ people.\label{fig:rgbm-mixing-time}}
\end{figure}

In \Fref{rgbm-mixing-relaxation-time} we display the estimated relaxation and mixing times as a function of the reallocation rate for various noise amplitudes. These numerical results indicate that the mixing times estimated with $\varepsilon = 1/4$ and relaxation times lead to similar conclusions about the extent of mobility across the whole disribution in RGBM. There is only slight dependence of the mixing time on the noise amplitude for $\tau\to 0$, but this dependence becomes insignificant very quickly. Therefore, we use the definition for mixing time as given by~\eqref{eq:mixing-time}.

\begin{figure}[!htb]
\centering
\includegraphics[width=12cm]{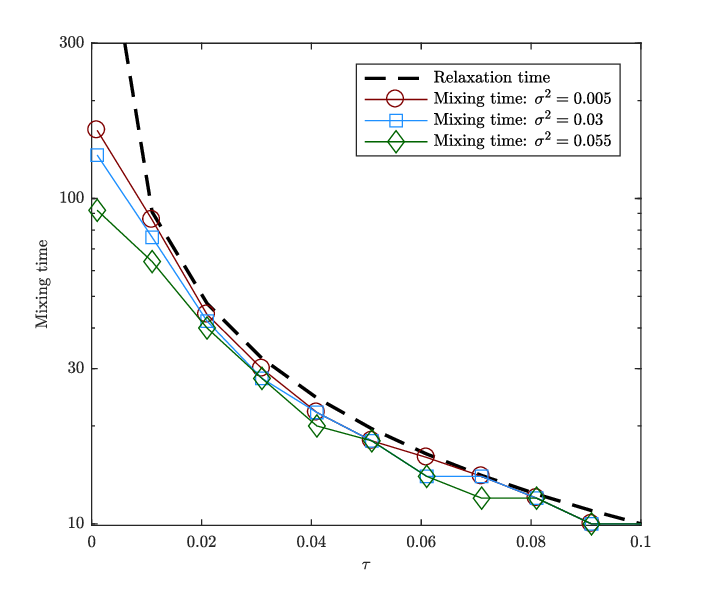}
\caption{\textbf{Mixing time and Relaxation time in RGBM.}  Mixing time in RGBM as a function of $\tau$ for different noise amplitudes with subsample size equal to 20\% of the population. In the simulations $N = 10^5$ people.\label{fig:rgbm-mixing-relaxation-time}}
\end{figure}


\section{Definitions of standard mobility measures}\label{sec:standard-mobility-measures}

\setcounter{equation}{0}
\setcounter{figure}{0}
\setcounter{table}{0}
\setcounter{subsection}{0}
\setcounter{subsubsection}{0}
\makeatletter
\renewcommand{\theequation}{A2.\arabic{equation}}
\renewcommand{\thesubsection}{A2.\arabic{subsection}}
\renewcommand{\thetable}{A2.\arabic{table}}
\renewcommand{\thefigure}{A2.\arabic{figure}}

\paragraph{Spearman's rank correlation:} The Spearman's rank correlation, $\rho_{t_0,t_f}$, is defined on a joint distribution of wealth ranks at two points in time, $t_0$ and $t_f$ ($t_0 < t_f$). It is defined as
\be
    \rho_{t_0,t_f} = 1 - \frac{6\sum_i \left[rg\left(\mathrm{x}_i\left(t_0\right)\right) - rg\left(\mathrm{x}_i\left(t_f\right)\right)\right]^2}{N\left(N^2-1\right)}\,,
\ee
where $rg(\mathrm{x})$ is the rank transformation of $\mathrm{x}$, $\mathrm{x}_i(t)$ is the wealth of individual $i$ in period $t$ and $N$ is the population size. This measure is bounded between $-1$ and $1$. $\rho_{t_0,t_f} = 1$ suggests perfect immobility, a state in which there is no change in wealth ranks between the two points in time. Lower values suggest greater economic mobility.

\paragraph{Intragenerational earnings elasticity:} The intragenerational earnings elasticity is defined as the slope $b_{t_0,t_f}$ of the regression
\be
   \log\left(\mathrm{x}_i\left(t_f\right)\right) = b_0 + b_{t_0,t_f} \log\left(\mathrm{x}_i\left(t_0\right)\right) + u_i\,,
\ee
where $b_0$ is the intercept and $u_i$ is the error term. This is a simple linear regression and therefore,
\be
    b_{t_0,t_f} = \mathrm{corr}\left(\log\left(\mathrm{x}\left(t_f\right)\right),\log\left(\mathrm{x}\left(t_0\right)\right)\right) \frac{\mathrm{var}\left(\log\left(\mathrm{x}\left(t_f\right)\right)\right)}{\mathrm{var}\left(\log\left(\mathrm{x}\left(t_0\right)\right)\right)}\,,
    \label{eq:iee-estimation}
\ee
where $\mathrm{corr}(\mathrm{x},\mathrm{y})$ is the correlation between the variables $\mathrm{x}$ and $\mathrm{y}$ and $\mathrm{var}(\mathrm{x})$ is the variance of $\mathrm{x}$. As with the rank correlation, lower IGE also indicates greater mobility. However, this measure is unbounded and may take on any real values.

\paragraph{Transition matrices \& Gumbel copula:} The wealth transition matrix disaggregates wealth rankings into quantiles and summarizes economic mobility in a transition matrix $\mathbf{A}$ in which the elements $A_{kl}$ quantify the probability that an individual in wealth quantile $k$ in period $t_0$ is found in wealth quantile $l$ in period $t_f$. In a perfectly mobile economy, the entries of the transition matrix are all equal to each other. This would correspond to $0$ rank correlation. In an immobile economy, on the other hand, the largest values are concentrated in the diagonal entries. A perfectly immobile case, of rank correlation $1$, would correspond to the identity transition matrix.\footnote{One might argue that mixing as a concept is equivalent to the notion of \textit{irreducibility} in transition matrices. However, transition matrices already aggregate the wealth dynamics into quantiles and thus may distort the picture of the extent of mobility.}

These matrices describe the bivariate joint wealth distribution at two points in time. Such distributions are usually modeled via copulas. Mathematically, a copula can be represented by a simple model in which the wealth transition matrix is parametrized. A widely used model is the Gumbel copula. It is able to reproduce realistic wealth-rank transition matrices, representing higher mobility at the bottom of the distribution than at the top \cite{JanttiJenkins2015}. The Gumbel copula is uniquely defined by a single parameter $\theta$, and as a result, this parameter can be treated as a measure of mobility where a larger value implies less mobility. 

\section{PSID data cleaning procedure}\label{sec:psid-data-cleaning}

\setcounter{equation}{0}
\setcounter{figure}{0}
\setcounter{table}{0}
\setcounter{subsection}{0}
\setcounter{subsubsection}{0}
\makeatletter
\renewcommand{\theequation}{A3.\arabic{equation}}
\renewcommand{\thesubsection}{A3.\arabic{subsection}}
\renewcommand{\thetable}{A3.\arabic{table}}
\renewcommand{\thefigure}{A3.\arabic{figure}}

The raw PSID dataset is available at \url{psidonline.isr.umich.edu}, but it needs to be cleaned for the purposes of our estimations. We clean the data as follows.

\begin{itemize}
    \item The dataset includes two different wealth variables: wealth and wealth without equity. The wealth variable is constituted of several asset types, net of debt value, and the value of home equity. The wealth without equity is constituted of the same data minus the home equity. In our estimations, we use the wealth variable as it includes all the possessions of a household.
    \item We estimate the target distribution using the data for 2019 and using the population weights that correspond to the same year.
    \item This dataset has longitudinal data on the socio-economic characteristics of 6000 US households between 1970 and 2019. Unfortunately, the data on some households in some of the surveys is missing. Therefore, we restrict our sample, used to create the subsample distribution, to the households that have data in every survey year between 1984 and 2019. These are in total 1143 households.
    \item To make the the resulting sample comparable to the population data, we re-weight the data as suggested by Perez et al.~\cite{perez2010re}.
    \item To create the potential subsample, we use data only on households in which the age of the head was between 30 and 35 years in 1984. This is because empirical observations have shown that age affects the ability of individuals to move across the wealth ladder~\cite{jianakoplos1997wealth,steckel2006wealth}. We choose households where the head is between 30 and 35 years of age because these are households representing relatively young families and as such should be more representative of mixing~\cite{sewell2003we,read2014social}.
    \item To estimate the subsample disribution we always use the weights corresponding to the initial time point. For example, if we estimate the mixing time with a starting point in 1984, we use the 1984 weights for creating the subsample distribution in each survey year after 1984. This makes the composition of the subsample the same and comparable across years. In our estimations, the weighted subsample size is about 20\% of the population.
\end{itemize}


\end{document}